\documentclass{emulateapj} %
\usepackage{graphicx}
\usepackage[usenames,dvips]{color}
\usepackage{color}
\usepackage{soul}

\def\arcsec{$\,^{\prime\prime}$~}
\def\arcmin{$\,^\prime$~}

\newcommand{\be}{\begin{equation}}
\newcommand{\bel}[1]{\begin{equation}\label{eq:#1}}
\newcommand{\ee}{\end{equation}}
\newcommand{\bd}{\begin{displaymath}} 
\newcommand{\ed}{\end{displaymath}}   
\newcommand{\bea}{\begin{eqnarray}}
\newcommand{\beal}[1]{\begin{eqnarray}\label{eq:#1}}
\newcommand{\eea}{\end{eqnarray}}

\newcommand{\eqref}[1]{\ref{eq:#1}}

\newcommand{\lsim }{{\lower0.8ex\hbox{$\buildrel <\over\sim$}}}
\newcommand{\gsim }{{\lower0.8ex\hbox{$\buildrel >\over\sim$}}}

\def\Chandra{${\it Chandra}$}
\def\HST{${\it HST}$\ }

\def\simge{\mathrel{%
   \rlap{\raise 0.511ex \hbox{$>$}}{\lower 0.511ex \hbox{$\sim$}}}}
\def\simle{\mathrel{
   \rlap{\raise 0.511ex \hbox{$<$}}{\lower 0.511ex \hbox{$\sim$}}}}

\newcommand{\Msun}{\ifmmode {M_{\odot}}\else${M_{\odot}}$\fi}
\newcommand{\Lsun}{\ifmmode {L_{\odot}}\else${L_{\odot}}$\fi}
\newcommand{\Rsun}{\ifmmode {R_{\odot}}\else${R_{\odot}}$\fi}

\shorttitle{Deeper Chandra Spectra of NGC 6652}
\shortauthors{Stacey et al.}

\begin{document}
\title{An Examination of the X-ray Sources in the Globular Cluster NGC 6652}  

\author{W.~S. Stacey\altaffilmark{1}, C.~O. Heinke\altaffilmark{1,2}, H.~N. Cohn\altaffilmark{3}, P.~M. Lugger\altaffilmark{3}, A. Bahramian\altaffilmark{1}}

\altaffiltext{1}{University of Alberta Physics Dept., CCIS 4-183, Edmonton, AB T6G 2E1, Canada}
\altaffiltext{2}{Ingenuity New Faculty; heinke@ualberta.ca}
\altaffiltext{3}{Department of Astronomy, Indiana University, 727 East 3rd St., Bloomington, IN 47405, USA}


\begin{abstract}
We observed the globular cluster NGC 6652 with \Chandra\ for 47.5 ks, detecting six known X-ray sources, as well as five previously undetected X-ray sources. Source A (XB 1832-330) is a well-known bright low-mass X-ray binary (LXMB).
The second brightest source, B, has a spectrum that fits well to either a power-law model ($\Gamma$ $\sim$ 1.3) or an absorbed hot gas emission model (kT $\sim$ 34 keV). Its unabsorbed 0.5-10 keV luminosity ($L_X$ = 1.6$^{+0.1}_{-0.1}\times10^{34}$ erg s$^{-1}$) is suggestive of a neutron star primary; however, Source B exhibits unusual variability for a LMXB, varying by over an order of magnitude on timescales of $\sim$ 100 s.
Source C's spectrum contains a strong low-energy component below $\sim$ 1 keV. Its spectrum is well fit to a simplified magnetic cataclysmic variable (CV) model, thus the soft component may be explained by a hot polar cap of a magnetic CV. 
Source D has an average $L_X$ (0.5-10 keV) $\sim$ 9$\times10^{32}$ erg s$^{-1}$ and its spectrum is well fit to a neutron star atmosphere model. This is indicative of a quiescent neutron star LXMB, suggesting Source D may be the third known LMXB in NGC 6652.
Source E has $L_X$ (0.5-10 keV) $\sim$ 3$\times10^{32}$ erg s$^{-1}$, while Source F has $L_X$ (0.5-10 keV) $\sim$ 1$\times10^{32}$ erg s$^{-1}$. Their relatively hard X-ray spectra are well-fit by power-law or plasma emission models.
Five newly detected fainter sources have luminosities between 1-5 $\times10^{31}$ erg s$^{-1}$.  NGC 6652 has an unusually flat X-ray luminosity function compared to other globular clusters, which may be connected to its extremely high central density.

\end{abstract}


\maketitle


\section{Introduction}\label{s:intro}

Low-luminosity X-ray sources in globular clusters 
were identified with Einstein \citep{Hertz83} as a separate class of X-ray sources from 
bright ($10^{36}$$ <$$ L_X$$ <$$ 10^{38}$ ergs/s) LMXBs containing neutron stars (NSs). 
They suggested that low-luminosity sources were primarily cataclysmic variables (CVs), 
with similar accretion rates to the LMXBs but potential energy wells 
1000 times shallower, but also including some NS LMXBs in quiescence (qLMXBs). 
\citet{Verbunt84} argued that the brightest low-$L_X$ systems 
($L_X$$>$$10^{33}$ ergs/s) were too bright to be CVs, and must be qLMXBs. 
The answer, from \Chandra\ observations, has been mixed; 
some of the brighter low-$L_X$ systems 
have been identified with qLMXBs, and some with CVs (e.g. in 47 Tuc, 
\citealt{Grindlay01}).  

The brighter low-$L_X$ 
systems ($5\times10^{32}$$ <$$ L_X$$ <$$ 10^{35}$ ergs/s) 
are of special interest.  
The bright CVs in this range are at the top of the X-ray luminosity 
distribution of CVs.  As disk-fed CVs transition to optically thick flows at high 
mass transfer rates, suppressing hard X-rays \citep{Patterson85},
 it is thought that reaching high X-ray luminosities may 
require magnetically channeled accretion (permitting optically thin flows) 
onto massive white dwarfs \citep[e.g.][]{Grindlay95, Ivanova06}.  The relative 
lack of dwarf novae outbursts in globular clusters also supports the magnetically 
channeled accretion idea \citep{Shara96}.  However, there is little \textit{direct} 
evidence for magnetic CVs in globular clusters; published evidence consists of 
$\lambda$4686 He II emission (associated with intermediate polar CVs, in which the accretion disk is truncated by the primary's B field) from 
3 CVs in NGC 6397 \citep{Edmonds99}, and very soft blackbody-like components 
in the CVs X10 in 47 Tuc \citep{Heinke05} and 1E 1339.8+2837 in M3 \citep{Stacey11}.
Further evidence of the nature of the most luminous CVs, in particular direct evidence 
of strong magnetic fields, will be important.

The most luminous qLMXBs ($10^{35}>L_X>10^{34}$ ergs/s) are brighter than the expected emission 
from residual heat in the crust \citep{Yakovlev04}, so must be powered by continuing accretion.  
 However, they are fainter than expected for X-ray 
binaries undergoing outbursts driven by the standard disk instability model 
\citep[e.g.][]{King00,Wijnands06}.  We don't yet understand the behavior of LMXBs that maintain 
these intermediate luminosities for years, such as XMMU J174716.1--281048 \citep{delSanto07} and M15 X-3 \citep{Heinke09}.
The neutron star's magnetic field is expected to exercise a ``propeller
 effect'' which should stop the accretion of material onto the neutron star 
surface at low mass transfer rates \citep{Illarionov75}.  
On the other hand, simulations of the interaction of weak accretion 
with a magnetic propeller predict some material will reach the 
neutron star \citep[e.g.][]{Romanova02}, suggesting that NS LMXBs in a propeller state might produce accretion at intermediate luminosities.

NGC 6652 contains a moderately bright ($L_X=10^{36}$ ergs/s) LMXB (A).  
Three other sources were identified in a 
1.6 ksec \Chandra\ HRC observation in 2001 \citep{Heinke01}. The brightest faint source, B, at $L_X$(0.5--7)$\sim6\times10^{33}$ ergs/s, has an optically (Hubble Space Telescope, HST) identified blue variable counterpart \citep{Deutsch98}. It is UV-bright but appears on the main sequence in a $V$ vs. $V-I$ color-magnitude diagram, suggesting a quiescent LMXB with a weak disk \citep{Heinke01}. 
A 5-ks 2008 \Chandra\ observation showed rapid variability, from $2\times10^{33}$ up to $>5\times10^{34}$ ergs/s on timescales $<$2 minutes, not previously seen from LMXBs at such low luminosities. Spectral fitting of countrate-selected bins suggested spectral variability, but it was unclear whether the behavior was true flaring or variations in obscuring column from an edge-on source \citep{Coomber11}.

The next brightest source, C, is very blue in $V-I$ HST observations, which combined with its $L_X$ of $6\times10^{32}$ ergs/s, indicates it is likely a very luminous CV \citep{Heinke01}. The 5-ks \Chandra\ observation found an extremely soft spectrum, with most of the 90 counts below 0.8 keV \citep{Coomber11}. D, at $L_X=5\times10^{32}$ ergs/s in the (optically crowded) cluster core, also showed a soft spectrum, with a suggestion of a line in the (73-count) spectrum.  The 5-ks \Chandra\ observation revealed three other faint sources, of which G lies outside our field of view.

NGC 6652 has been considered to be a cluster of only moderate central density \citep{Harris10} and stellar encounter rate \citep{Verbunt02}, making its possession of a bright LMXB somewhat  unusual.  However, a recent surface brightness profile from \citet{Noyola06} finds a sharp core with a significantly higher central surface brightness than previously measured, identifying it as a core-collapsed cluster.  We have computed the central luminosity densities of Galactic globular clusters using the Noyola \& Gebhardt parameter updates, and note that NGC 6652's central density is the fifth highest.

\section{Data Reduction}
We observed the globular cluster NGC 6652 on June 3rd, 2011 with the $\it{Chandra}$ ACIS-S detector, using two active CCDs (S2 and S3) with a 1/8 subarray to abate pileup from Source B. The total observation time was 47.5 ks.
The data were reduced using CIAO version 4.3\footnote{http://cxc.cfa.harvard.edu/ciao/}. A new bad pixel file was generated with the CIAO $\it{acis\_run\_hotpix}$ tool. 
The observations were further processed using $\it{acis\_process\_events}$ and filtered\footnote{http://cxc.cfa.harvard.edu/ciao/threads/createL2/} in order to create level 2 event files, as detailed in the CIAO Science Threads\footnote{http://cxc.cfa.harvard.edu/ciao/threads/all.html}.

\begin{figure}
\figurenum{1}
\includegraphics[scale=0.5]{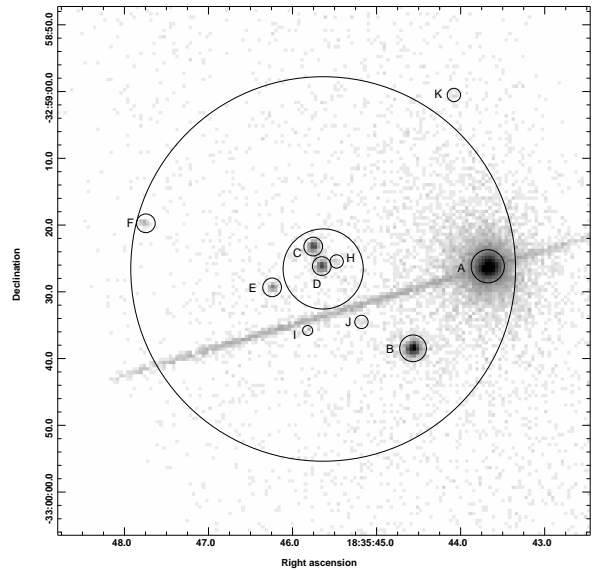}
\caption[fig_xrayimage.ps]{ \label{fig:fig1}
 ACIS-S image of NGC 6652 over 0.3--7 keV, with X-ray sources are labeled. The core radius (0.10\arcmin) and half-mass radius (0.48\arcmin) are shown \citep{Harris10}. A previously detected  X-ray source (Source G) lies outside our field of view (see \citet{Coomber11}), while Source L lies $\sim$1.4\arcmin from the center of the cluster.
} 
\end{figure}

We ran the CIAO $\it{wavdetect}$ tool \footnote{http://cxc.harvard.edu/ciao/threads/wavdetect/} over an energy range of 0.5--10 keV. From this, we found six X-ray sources, all located within the cluster half-mass radius (Figure \ref{fig:fig1}). A seventh previously detected source \citep{Coomber11} lay outside our field of view. 
The source positions agree with our previous 2008 $\it{Chandra}$ observation.
In addition, we ran $\it{wavdetect}$ over several other ranges (0.3--7 keV, 0.3--2 keV, 2--7 keV); however, no previously undetected sources were found. We further ran $\it{PWDetect}$\footnote{http://www.astropa.unipa.it/progetti\_ricerca/PWDetect/} \citep{Damiani97i,Damiani97ii} over a range of 0.3--7 keV. Both $\it{PWDetect}$ and $\it{wavdetect}$ use wavelet transforms to locate possible sources; we have found that $\it{wavdetect}$ efficiently finds sources without spurious detections over large uncrowded fields of view, while $\it{PWDetect}$ can more effectively identify faint sources near bright sources, though it has a higher spurious detection rate, making it less suitable for large fields \citep[e.g.][]{Heinke03c,Bogdanov10a}.  
In order to reduce the number of spurious detections, we only include detections above 5 $\sigma$. We identify four previously undetected X-ray sources within the cluster half-mass radius (Sources H through K) and one source outside the cluster half-mass radius (Source L).  
Our detection limit of 10 counts translates to roughly $1\times10^{31}$ ergs/s (0.5--2.5 keV), or $2\times10^{31}$ ergs/s (0.5--10 keV) for standard absorbed powerlaw spectra with photon index=2 at the 10 kpc cluster distance \citep{Harris10}.
  
We selected events within a circle around the sources (radius 1\arcsec for H, J, K and L, radius 0.75\arcsec for I, due to its proximity to the readout streak), and background from local regions, determining the background subtracted source counts. The positions and net counts for each observed source are listed in Table \ref{tab:positions}, with 95\% confidence positional uncertainty values from equation (5) of \citet{Hong05}.
For Sources B through F, we selected events from a circular source region (2\arcsec for B, 1.4\arcsec for C through F) centered on the $\it{wavdetect}$ positions. For Sources B, E and F, an annulus of inner radius 5\arcsec that excluded the source was chosen for the background. Due to the proximity of Sources C and D, a larger annulus of inner radius 6\arcsec that excluded both sources was chosen for the background. Using these regions, the CIAO $\it{specextract}$ script was run to extract source and background spectra, generating RMF and ARF files. XSPEC version 12.7\footnote{http://heasarc.gsfc.nasa.gov/docs/xanadu/xspec/} was used to examine and fit the spectra (discussed in the following sections). 

\section{X-ray Analysis}
\subsection{Source A}
We extracted 1537 counts from Source A's readout streak (using a 99.5 by 3.5 pixel box region), grouping them into 14 bins with a minimum of 100 counts per bin.  We were able to successfully ($\chi^2_{\nu}$=1.05 for 11 degrees of freedom) fit Source A's spectrum to an absorbed power-law model ($\Gamma$ = 1.9 $\pm$ 0.3), with $L_X$(0.5--10 keV)=7.0$^{+0.7}_{-0.7}\times10^{35}$ ergs/s. A hydrogen column density of 27$^{+11}_{-9}\times10^{20}$ cm$^{-2}$ was measured, significantly larger than the cluster value of 5$\times10^{20}$ cm$^{-2}$ \citep{Harris10}.  Previous studies of A's X-ray spectrum with ASCA, BeppoSAX, and XMM \citep{Mukai00,Parmar01,Sidoli08} agree in requiring additional intrinsic absorption, with their single-component best-fit spectra resembling ours.  Those previous observations (optimized for this bright source) were able to better constrain A's spectrum, measuring partial covering and multiple spectral components, which our data are insufficient to constrain.  

\begin{figure}
\figurenum{2}
\includegraphics[scale=0.45]{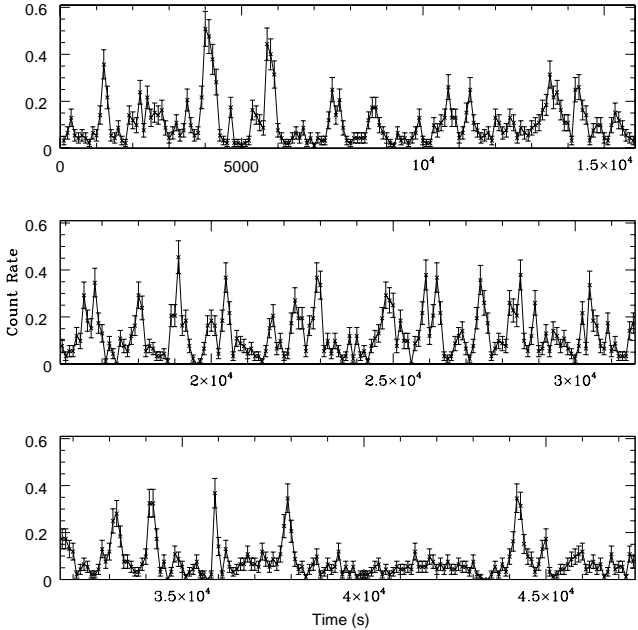}
\caption[fig_lightcurveb.ps]{ \label{fig:fig2}
 ACIS lightcurve extracted from Source B in the energy band 0.3--7 keV, using a binning of 100 s. The lightcurve is split into three consecutive periods (top, middle and bottom) in order to clearly display the peaks in the lightcurve. A corresponding variability index of 10 (variable) was found using the CIAO $\it{glvary}$ tool.
} 
\end{figure}

\subsection{Source B}
\subsubsection{Timing Analysis}\label{s:timeb}
A 474 bin lightcurve (100 s/bin) was extracted for Source B using the CIAO $\it{dmextract}$ tool (Fig. \ref{fig:fig2}). The observation shows clear variability by factors over an order of magnitude on timescales of $\sim100$ s, uncharacteristic of low-mass X-ray binaries. We verified the variability with the CIAO $\it{glvary}$ tool, which applies the Gregory-Loredo algorithm. The algorithm separates the observation into bins based upon time and looks for significant variation, returning a variability index \citep{Gregory92}.  For Source B, a variability index of 10 was found, corresponding to a variability probability of ${\sim1.0}$.  Power spectra (generated using XRONOS\footnote{http://heasarc.nasa.gov/xanadu/xronos/xronos.html}) show no clear periodicity in Source B's X-ray emission (using either 100 s or 0.54 s binning). In order to test whether the variation may be due to changes in the obscuring column, the observation was filtered into eight separate ranges based upon the count rate. The count rate ranges were chosen to ensure each range contains at least 400 counts. The hardness ratio within each range was then quantified by examining the ratio of the number of counts with photon energy between 0.3--1 keV to the number of counts with photon energy between 0.3--7 keV (see Table \ref{tab:specfitsb}). If the variation was due to obscuration, we would expect to find the lowest hardness ratios in the lowest count rate range. This is seen to some degree as the lowest hardness ratio (0.16$\pm0.02$) is found between 0.05-0.1 counts s$^{-1}$ and the largest hardness ratio (0.23$\pm0.02$) is seen between 0.27-0.35 counts s$^{-1}$. We further examined the spectrum for each of these ranges (below).

\subsubsection{X-ray Spectral Analysis}
Source B's total spectrum was binned (using the GRPPHA tool) so that each bin contained a minimum of 100 counts. We included a photoelectric absorption component (XSPEC model $\it{phabs}$) with each model, leaving the hydrogen column denstiy, $N_H$, as a free parameter. The metal abundance for the $\it{mekal}$ model was kept at [Fe/H] = -0.81 \citep{Harris96}\footnote{Updated 2010; http://www.physics.mcmaster.ca/$\sim$harris/mwgc.dat}. Considering a range from 0.5 to 10 keV, Source B's spectrum was fit reasonably well to both an absorbed power-law model (XSPEC model $\it{pegpwrlw}$, photon index $\Gamma=1.3\pm0.1$) and an absorbed hot X-ray plasma model (XSPEC model $\it{mekal}$, temperature kT=34$^{+28}_{-13}$ keV). 
We found an unabsorbed X-ray flux (0.5--10 keV) of (1.4$\pm0.1)\times10^{-12}$ erg cm$^{-2}$ s$^{-1}$ from the power-law model and (1.3$\pm0.1)\times10^{-12}$ erg cm$^{-2}$ s$^{-1}$ from the plasma emission model. Using a distance of 10.0 kpc \citep{Harris96}, these correspond to luminosities of (1.7$\pm0.1)\times10^{34}$ erg s$^{-1}$ and (1.6$\pm0.1)\times10^{34}$ erg s$^{-1}$, consistent with the 2008 $\it{Chandra}$ ACIS-S observation.
The full results are listed in Table \ref{tab:specfits}.  

\begin{figure}
\figurenum{3}
\includegraphics[scale=0.37, angle=270]{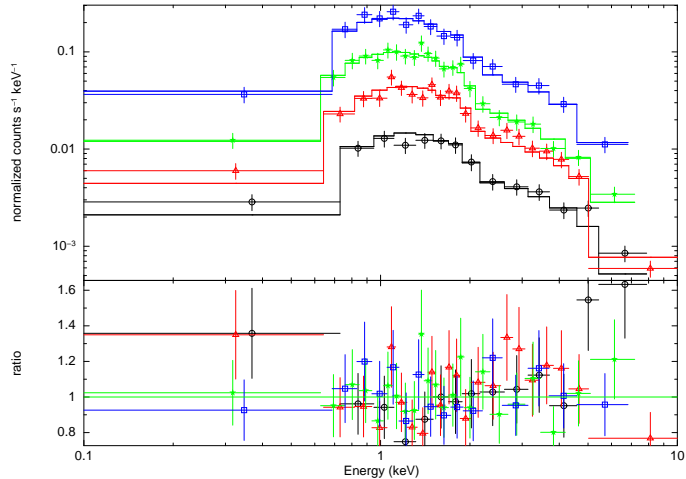}
\caption[fig_spectrum_sourceb2.ps]{ \label{fig:fig3}
X-ray spectra of Source B, with a minimum of 30 counts per bin. Spectra are filtered by count rate ranges, with best-fit absorbed $\it{mekal}$ models fit from 0.5 to 10 keV. Eight separate ranges were fit to produce the results in Table \ref{tab:specfitsb} but for clarity only four are shown here. From highest to lowest count rate, these are $>$0.35 counts s$^{-1}$ (blue, squares), 0.15 - 0.2 counts s$^{-1}$ (green, stars), 0.075 - 0.1 counts s$^{-1}$ (red, triangles) and $<$0.05 counts s$^{-1}$ (black, circles).  The ratio of the data to the model is shown below. 
} 
\end{figure}

We next filtered the spectrum into eight separate ranges based upon count rate in order to look for possible changes in the spectra. Using a minimum of 30 counts per bin, we fit the spectra to an absorbed $\it{mekal}$ model over 0.5 to 10 keV (Figure \ref{fig:fig3}). In addition, the five spectra with the largest count rates contained a pileup component (grade morphing parameter fixed at 0.5, \citealt{Davis01}). The X-ray plasma emission temperature and normalization, as well as the hydrogen column density ($N_H$), were allowed to vary. The parameters from these fits are listed in Table \ref{tab:specfitsb}. There appears to be a tendency for the higher count rate spectra to have lower $N_H$ values, as the highest $N_H$ observed appears in the 0.05-0.075 counts s$^{-1}$ range and the lowest appears in the 0.27-0.35 counts s$^{-1}$ range. However, this tendency is not strong; the maximum fitted $N_H$ is only $2.7^{+1.5}_{-1.4}\times10^{21}$ cm$^{-2}$, insufficient to significantly decrease the  countrate. Therefore, we rule out that the flux variations are entirely due to obscuration by changing $N_H$ columns, e.g. due to an edge-on accretion disk.

\begin{figure}
\figurenum{4}
\includegraphics[scale=0.45]{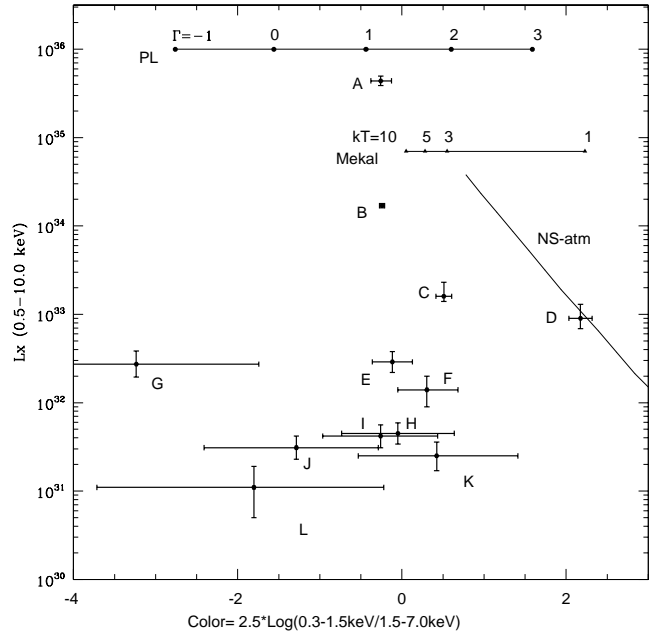}
\caption[fig_spectrum_sourcee.ps]{ \label{fig:fig9}
 X-ray color-magnitude diagram for detected sources. Color is determined as a function of low-energy to high energy counts. X-ray luminosity (0.5--10 keV) is plotted along the vertical axis. Errors are derived for 1 $\sigma$ from \citet{Gehrels86}. Values for Sources A and G are taken from \citet{Coomber11}. Additionally, the theoretical colors for absorbed power-law, $\it{mekal}$ and $\it{nsatmos}$ models (with the corresponding $L_X$ for {\it nsatmos}) are shown.
} 
\end{figure}

We placed Source B (as well as the other observed sources) on an X-ray color-magnitude diagram (Figure \ref{fig:fig9}). We defined color to be 2.5 log[(0.5-1.5 keV counts)/(1.5-7.0 keV counts)] and plotted it against the 0.5--10 keV luminosity of the sources. In addition, we plotted the X-ray color predictions for power-law and $\it{mekal}$ models, as well as the X-ray color-luminosity relation for the $\it{nsatmos}$ model.

\subsection{Source C}
\begin{figure}
\figurenum{5}
\includegraphics[scale=0.43]{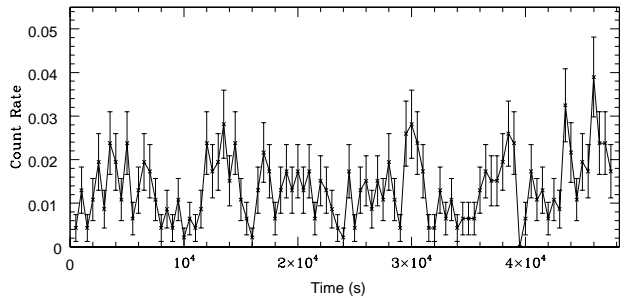}
\caption[fig_lightcurvec2.ps]{ \label{fig:fig4}
ACIS lightcurve extracted from Source C over 0.3--7 keV, using a binning of 500 s. The CIAO $\it{glvary}$ tool gave a variability index of 7 (variable).
} 
\end{figure}

A lightcurve (500 s/bin) was extracted for Source C (see Figure \ref{fig:fig4}). Using the CIAO $\it{glvary}$ tool (discussed in Section \ref{s:timeb}) a variability index of 7 was found, with a corresponding variability probability of 0.995. Power spectra show no clear evidence of periodicity. The total unabsorbed luminosity from the simplified polar CV model (below) is $L_X$ (0.5--10 keV) = 1.6$^{+0.7}_{-0.2}\times10^{33}$ erg s$^{-1}$, which is comparable to the 2008 ACIS-S observation which found $L_X$ (0.5-10 keV) = 1.2$^{+0.1}_{-0.5}\times10^{33}$ erg s$^{-1}$ \citep{Coomber11}.

\begin{figure}
\figurenum{6}
\includegraphics[scale=0.37, angle=270]{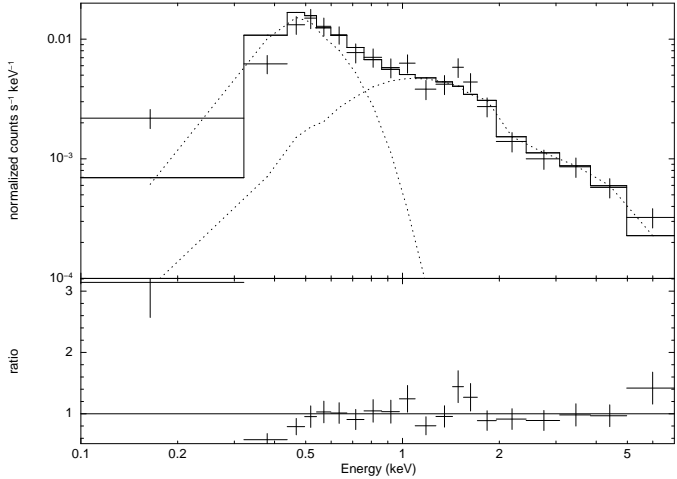}
\caption[fig_spectrum_sourcec2.ps]{ \label{fig:fig5}
$\it{Chandra}$ spectrum of Source C fit to an absorbed $\it{mekal}$ plus blackbody model from 0.5 to 10 keV, with a minimum of 30 counts per bin. The dotted lines indicate the two components of the model ($\it{mekal}$ being of higher energy). The ratio of the data to the model is shown below. Resulting parameters from the fit are listed in Table \ref{tab:specfits}. 
} 
\end{figure}

Grouping Source C's spectrum into bins with a minimum of 30 counts each, a notable low energy component below $\sim1$ keV was found (Figure \ref{fig:fig5}). 
At energies below $\sim0.5$ keV, the accuracy of the $\it{Chandra}$ response files and the sensitivity of the ACIS-S detector are no longer optimal. For this reason, we only considered a range from 0.5 to 10 keV for our primary fits.
Even so, we were unable to adequately model Source C to any single component model (see Table \ref{tab:specfits}). 
Allowing the $N_H$ to vary, we successfully fit the spectrum to a model containing two hot X-ray plasma emission components (XSPEC model $\it{mekal}$) with temperatures $>$17 keV and 0.18$\pm0.03$ keV. 
Double $\it{mekal}$ models are commonly used to describe the spectra of CVs \citep{Baskill05}; however, Source C's lower-energy component is unusually strong.
In order to examine the lower energy emission, we fit the same double $\it{mekal}$ model over 0.2 to 10 keV; however, this failed to produce an acceptable fit ($\chi^2_{\nu}$ = 1.99).
Source C's spectrum was adequately fit to a simplified polar CV model, containing a high-temperature $\it{mekal}$ component and a soft blackbody component (e.g. \citealt{Ramsay04}). Again, allowing the $N_H$ to vary, the best fit model resulted in a $\it{mekal}$ temperature $>$18 keV and a blackbody temperature of 70$\pm20$ eV.
Fitting down to 0.2 keV with this model gave a better fit than the double mekal model in the same range ($\chi^2_{\nu}$ = 0.95, $\it{mekal}$ temperature $>$ 14 keV, blackbody temperature of 80$\pm10$ eV).  

\subsection{Source D}
The lightcurve from Source D showed no evidence of variation. A variability index (discussed in Section \ref{s:timeb}) of 0 was found, resulting in a low variability probability of 0.033.

\begin{figure}
\figurenum{7}
\includegraphics[scale=0.37, angle=270]{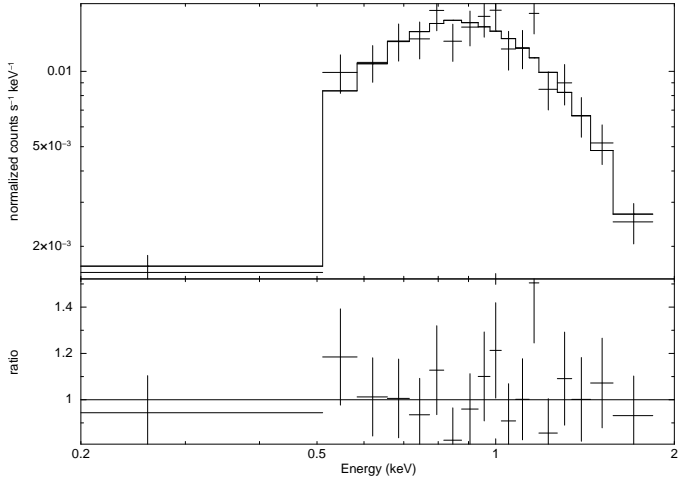}
\caption[fig_spectrum_sourced.ps]{ \label{fig:fig6}
$\it{Chandra}$ spectrum of Source D fit to an absorbed neutron star atmosphere model from 0.5 to 10 keV, with a minimum of 30 counts per bin. The ratio of the data to the model is shown below. Resulting parameters from the fit are listed in Table \ref{tab:specfits}. 
} 
\end{figure}

Source D's spectrum was extracted and grouped to ensure a minimum of 30 counts per bin (Figure \ref{fig:fig6}). Including a photon absorption component and allowing the $N_H$ to vary, Source D was fit to a power-law model resulting in a photon index of 5.8$^{+0.8}_{-1.2}$ (see Table \ref{tab:specfits}). The fitted $N_H$ of (42$^{+14}_{-19}$) $\times$10$^{20}$ cm$^{-2}$ is much larger than the cluster value of 5 $\times$10$^{20}$ cm$^{-2}$, and the high power-law photon index suggests testing thermal emission models. A $\it{mekal}$ model proved to be a poor fit; however, a neutron star atmosphere model (XSPEC model $\it{nsatmos}$, 1.4 \Msun, radius of 10 km) proved to sufficiently fit the spectrum. From this we found an effective temperature of 111$\pm2$ eV. 
We used the XSPEC $\it{cflux}$ model to find $L_X$ (0.5--10 keV) = 9.0$^{+4.0}_{-2.1}\times10^{32}$ erg s$^{-1}$, consistent with the 2000 $\it{Chandra}$ HRC observation and the 2008 $\it{Chandra}$ ACIS observation. 
Our previous $\it{Chandra}$ observation \citep{Coomber11} suggested the existence of an emission line near 1 keV; however, our higher-quality spectrum shows no evidence for this.

\subsection{Sources E + F}
We detected 99 counts from Source E and 35 counts from Source F. The relatively low number of counts reduces the accuracy of timing analysis; however, it is still possible to use the CIAO $\it{glvary}$ tool to look for strong variation. A variability probability of 0.78 was found for Source E (variability index of 2) and a variability probability of 0.45 was found for Source F (variability index of 0), thus we do not have strong evidence of variability in either.

\begin{figure}
\figurenum{8}
\includegraphics[scale=0.37, angle=270]{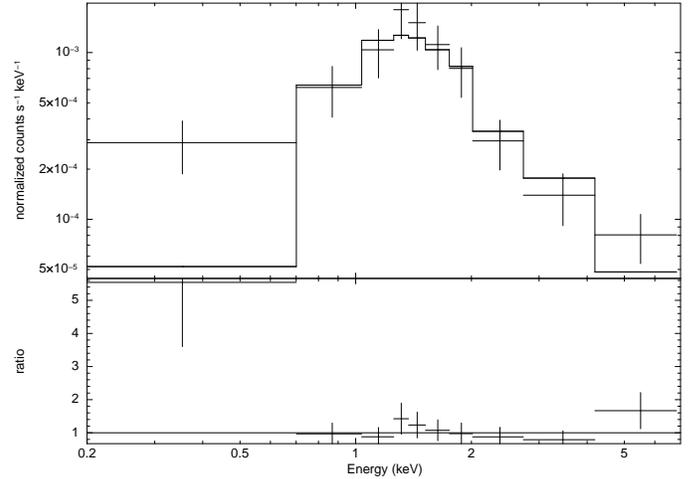}
\caption[fig_spectrum_sourcee.ps]{ \label{fig:fig7}
$\it{Chandra}$ spectrum of Source E fit to an absorbed power-law model over 0.5 to 10 keV, with a minimum of 10 counts per bin. The ratio of the data to the model is shown below. Resulting parameters from the fit are listed in Table \ref{tab:specfits}. 
} 
\end{figure}

The spectrum for Source E was extracted and grouped to ensure a minimum of 10 counts per bin (see Figure \ref{fig:fig7}). To ensure optimal accuracy from the $\it{Chandra}$ response files, we considered an energy range from 0.5 to 10 keV. Allowing the $N_H$ to vary, Source E was successfully fit to both an absorbed power-law model with photon index of 2.2$^{+1.1}_{-0.8}$ and an absorbed $\it{mekal}$ model with temperature $>$1.8 keV (Table \ref{tab:specfits}). 
An unabsorbed 0.5--10 keV luminosity of 2.9$^{+0.9}_{-0.7}\times10^{32}$ erg s$^{-1}$ was determined from the $\it{mekal}$ model, consistent with our previous 2008 $\it{Chandra}$ observation.

\begin{figure}
\figurenum{9}
\includegraphics[scale=0.37, angle=270]{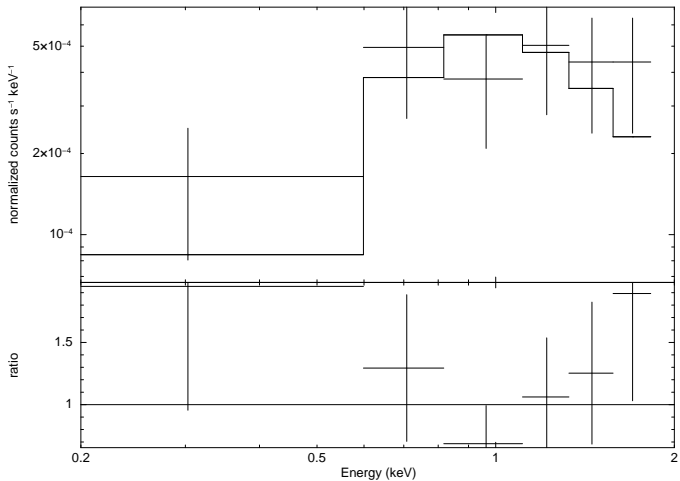}
\caption[fig_spectrum_sourcee.ps]{ \label{fig:fig8}
$\it{Chandra}$ spectrum of Source F. The model was determined using the C-stastistic, fitting an absorbed $\it{mekal}$ model to the unbinned data over 0 to 10 keV. For clarity only, the spectrum is shown with a minimum of 5 counts per bin. The ratio of the data to the model is shown below. Resulting parameters from the fit are listed in Table \ref{tab:specfits}. 
} 
\end{figure}

The spectrum for Source F was also grouped to ensure a minimum of 10 counts per bin, resulting in a total of four bins (Figure \ref{fig:fig8}). Due to the low number of bins, the entire 0-10 keV range was considered. We were unable to find an ideal single-component model with which to fit Source F. An absorbed $\it{nsatmos}$ model proved a poor fit, while the null hypothesis probability (NHP) for both an absorbed power-law ($\Gamma = 1.7^{+1.1}_{-0.5}$, NHP = 0.31) and an absorbed $\it{mekal}$ model (temperature $>$ 1.9 keV, NHP = 0.29) remained high. Using the C-statistic \citep{Cash79}, we further examined the suitability of each model by fitting the unbinned data. For the power-law model 99.9\% of simulated spectra resulted in lower C-statistic values, while 99.7\% did for the $\it{mekal}$ model and 100\% did for the $\it{nsatmos}$ model (for 10$^4$ simulated spectra), suggesting a poor fit in each case. The power-law and $\it{mekal}$ parameters for the unbinned fitting agree within error to the binned fitting parameters ($\Gamma$ $<$ 6.0, kT = 1.8$^{+2.8}_{-0.5}$). The unbinned $\it{nsatmos}$ model results in a marginally lower temperature of 85$^{+10}_{-8}$ eV. The luminosity obtained for Source F (Table \ref{tab:specfits}) was consistent with the previous 2008 $\it{Chandra}$ observation.

\subsection{Faint Sources}\label{sec:faint}
The five faintest sources have too few counts for accurate timing or spectral analysis. Using the number of source counts found for each source and the $\it{Chandra}$ PIMMS version 4.2, we estimated unabsorbed X-ray luminosities (0.5--10 keV) using a photon index of 1.4 (Table \ref{tab:positions}). There exists marginal evidence for a sixth faint source located between C and D; however, it was not identified by either $\it{wavdetect}$ or $\it{PWDetect}$ (possibly due to its proximity to C and D).  It would be difficult for more than one or two X-ray sources over 10 counts (translating to $L_X$(0.5--2.5 keV)$>10^{31}$ ergs/s) to be missed, apart from near source A or within 2'' of source D. 

\subsection{Radial Distribution}
\label{sec:RD}
We have analyzed the radial distribution of the cluster sources by
fitting generalized King model profiles using maximum likelihood
techniques (see, e.g. \citet{Grindlay02, Heinke06b, Cohn10}).
 In this approach, the projected radial distribution is
given by the expression,
\begin{equation}
\label{SB_profile}
 S(r) = S_0 \left[ 1 + \left({r\over r_0}\right)^2 \right]^{\alpha/2},
\end{equation}
where $\alpha$ is the large-$r$ power-law slope and $r_0$ is a radial
scale factor that is related to the core radius by $r_c =
(2^{-2/\alpha}-1)^{1/2}r_0$.  We assume that the turnoff mass
stellar distribution, which dominates the optical surface brightness
profile, is described by a King model, i.e.\ by Eqn.~\ref{SB_profile}
with $\alpha = -2$ and $r_c = r_0$.  In thermal equilibrium, the
distribution of the {\em Chandra} sources is expected to be described
by Eqn.~\ref{SB_profile} with $\alpha_X = -3q+1$, where $q = M_X/M_*$
is the ratio of the characteristic source mass to the mass of the
stars that dominate the optical surface brightness profile.  We here
adopted the values of $r_c = 5.8''$ and $r_h = 29''$ from \citet{McLaughlin05}
 for the optical surface brightness profile.
It is first necessary to correct the source distribution for
background contamination.  Based on the \citet{Giacconi01}
extragalactic source counts, we estimate a total of 2.1 background
sources within the $1'\times8'$ subarray.  We note that there is one
source (L) that lies well outside of the half-mass radius, which is
roughly consistent with this predicted level.  This background level
implies approximately 0.2 background sources within $r_h$.  This in
turn suggests that essentially all of the detected sources within
$r_h$ belong to the cluster.  We note that we have not made a
correction for any incompleteness due to fainter sources being lost in
the vicinity of the very bright source A.

\begin{figure}
\figurenum{10}
\includegraphics[scale=0.35]{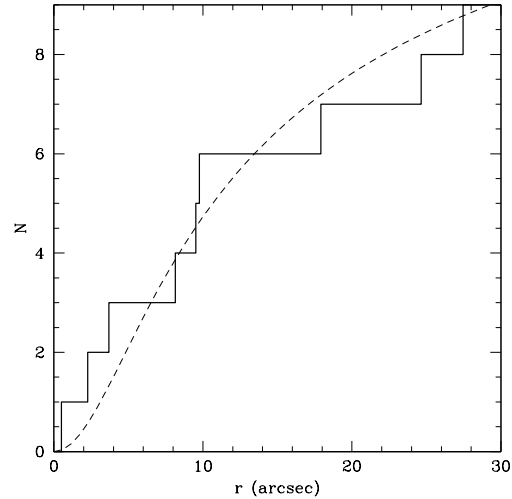}
\caption[fig_radial.ps]{ \label{fig:fig10}
Cumulative radial distribution of {\em Chandra} sources in
NGC 6652 (solid line) with a generalized King model fit (dashed line).
The sample includes the nine sources within the half-mass radius.
} 
\end{figure}

Given the small source sample, we have carried out one-parameter fits
by finding the value of $q$ which maximizes the likelihood.  As in our
previous work, we use bootstrap resampling to estimate the best fit
parameter uncertainties.  The parameters $r_{c,X}$ and $\alpha_X$ are
expressed as functions of $q$, and $S_0$ is determined by
normalization.  The resulting maximum likelihood value of the mass
ratio is $q = 1.17 \pm 0.28$.  The corresponding core radius and slope
values are $r_{c,X} = 5.0 \pm 0.8$ and $\alpha_{c,X} = -2.5 \pm 1.0$.
This fit to the cumulative radial source distribution is shown in
Fig. \ref{fig:fig10}.  For an assumed main-sequence turnoff mass (MSTO) of $0.80
\pm 0.05~M_\odot$, the corresponding characteristic mass for the
Chandra sources is $0.94 \pm 0.23~M_\odot$.  While this provides a
hint that $M_X$ exceeds the MSTO mass, the difference is clearly not
statistically significant.  The large statistical uncertainty is
likely due to the small sample size.

We note that \citet{Noyola06} have classified NGC 6652 as
core collapsed, with an optical core radius of $1.2''$.  As we did in
our previous analysis of the {\em Chandra} source distribution in M30
\citep{Lugger07}, we have fitted the radial distribution of sources with
a pure power law for comparison with the optical surface brightness
profile.  The resulting slope is $\alpha_X = -1.30 \pm 0.22$.  This is
significantly steeper than the slope of $-0.57 \pm 0.12$ measured by
 \citet{Noyola06} for the optical cusp.  In principle,
comparison of these slopes can be used to estimate $M_X$.  However,
the slope of the profile presented in their paper continuously
steepens with increasing radius beyond the core, rather than
maintaining a constant value for a significant run of radius as in
other core collapsed clusters.  The characteristic slope at $10''$ is
about $-1.3$, which is the same as the slope for the {\em Chandra}
sources.  This suggests a $q$ value of unity, putting $M_X$ near the
MSTO mass.  Once again, there is considerable statistical uncertainty
in this value.  In sum, the spatial distribution of sources does not
provide a definitive estimate of the source mass.

\section{Discussion}

\subsection{Source A}
Source A's unabsorbed X-ray luminosity is $L_X$ (0.5--10 keV) = 7.0$^{+0.7}_{-0.7}\times10^{35}$ ergs/s. This is comparable to our 2008 observation which found Source A to have a luminosity of $L_X$ (0.5--6.0 keV) = 4.4$^{+0.6}_{-0.5}\times10^{35}$ ergs/s (\citealt{Coomber11} ; our 0.5--6.0 keV luminosity is 5.4$^{+0.}_{-0.4}\times10^{35}$ ergs/s). Observations of XB 1832--330 previous to our 2008 observation \citep{Sidoli08, Tarana07, Parmar01} have reported X-ray luminosities several times higher.  We note that the RXTE bulge scan monitoring observations \citep{Swank01} \footnote{ http://asd.gsfc.nasa.gov/Craig.Markwardt//galscan/main.html} show a roughly sixfold drop in flux in early 2011.  Our observation suggests that this decline in A's luminosity has not continued smoothly.

The scheme for dividing LMXBs into ultracompact (white dwarf donors) vs. long-period systems based on X-ray spectra \citep{Sidoli01} suggested Source A (XB 1832-330) as ultracompact \citep{Parmar01}, which is ruled out by the 2.15 hour period measurement of \citet{Engel12}.  This scheme may still be useful, but in dividing short-period ($\lsim$2 hours) from long-period ($\gsim$6 hours) systems.

\subsection{Source B}
The peak luminosity of Source B suggests that the system contains a neutron star; however, Source B's extreme variability is unusual for a LMXB. In May 2011, a 6.5 hour observation of NGC 6652 taken with Gemini's GMOS-S CCD detector using g' and r' filters found a similar, highly variable lightcurve for Source B \citep{Engel12}. No clear periodicity was found.
 
One previously suggested explanation for Source B's behavior is a system with a high inclination angle, so that our view of the central X-ray object is obscured by the accretion disk. If this were the case, we would expect Source B's spectrum to be hardened during periods of low count rate. We see marginal evidence for this as the hardness ratio ranges from 0.16$\pm0.02$ for the second-lowest count rate range to 0.23$\pm0.02$ for the second-highest count rate range. While the tendency is not clear, the $N_H$ does appear to decrease with increasing count rate. However, the maximum $N_H$ ($2.7^{+1.5}_{-1.4}\times10^{21}$ cm$^{-2}$) is not sufficient to significantly decrease the countrate. An edge-on accretion disk, where flux variations are largely  due to obscuration by changing $N_H$ columns, can therefore be ruled out.

The behavior of Source B may be better explained by an instability involving the propeller effect, when the magnetosphere of the neutron star rotates faster than the innermost region of the accretion disk, pushing material away \citep{Illarionov75}. The location of the inner edge of the accretion disk will be determined by the ratio of gas to magnetic pressure; as gas accumulates in the disk, the disk will come closer, enabling accretion.  This can produce cyclic mass transfer on a range of timescales, which have been explored in simulations by several theorists \citep{Spruit93,D'Angelo10,Romanova11}.
Such an instability has been suggested to explain $\sim$ 1 Hz modulation in the transient LMXB SAX J1808.4--3658 \citep{Patruno09}, and the behavior of the 'Rapid Burster' \citep{Spruit93}. A similar phenomenon may be responsible for the rapid flaring in Source B.

Source B has the properties of a very faint X-ray transient (VFXT). These systems typically have neutron star or black hole primaries and have peak X-ray luminosities $10^{34} < L_X < 10^{36}$ with quiescent $L_X$ generally an order of magnitude lower \citep{Wijnands06, Muno05b}. 
Their time-averaged mass transfer rates are rather lower than expected for most evolutionary states.  One explanation is that the majority of the inflowing gas is ejected, through e.g. a propeller effect.  Weak and strong propeller regimes, corresponding to weak and strong outflows, have been studied by MHD simulations \citep{Romanova05, Ustyugova06}. 
 If the propeller effect is truly responsible for the characteristics of Source B, it may prove important to understanding the nature of other VFXTs.

\subsection{Source C}
The soft spectral component in Source C's spectrum is consistent with some magnetic CV spectra. Source C fits well to a simplified polar CV model, where a higher energy $\it{mekal}$ component describes hard X-ray emission originating the accretion shock column front, while a lower energy blackbody component describes soft X-ray emission from the white dwarf's polar caps \citep{Ramsay04}. While Source C is several times more luminous than the suggested polar CV X10 in 47 Tuc \citep{Heinke05}, its spectrum is comparable. Source C may therefore be the second polar CV identified in a globular cluster.

\subsection{Source D}
Source D's spectrum was fit well to a neutron star atmosphere model, while a $\it{mekal}$ model proved to be an exceptionally poor fit. We found Source D to have low variability over our observation, while its inferred luminosity is comparable to that seen in the 2008 observation. Source D's luminosity is similar to other known quiescent LMXBs in globular clusters \citep{Heinke03d}. As well, it lies close to the theoretical cooling track for a neutron star atmosphere on the color-magnitude diagram. We therefore suggest that Source D is a NS qLMXB.

\subsection{Fainter Sources}
The brightest of the fainter sources, Source E's spectrum and luminosity suggest that it may be a CV. However, due to the low number of counts, the evidence is not conclusive. We were unable to adequately model Source F with any single component model, though a neutron star atmosphere model proved an exceptionally poor fit. 
Sources H through K cannot be conclusively classified. However, most X-ray sources in globular clusters with similar luminosities and X-ray colors are CVs \citep{Pooley06}, with smaller fractions of chromospherically active binaries and millisecond pulsars (MSPs); therefore we may suggest the majority of our fainter sources are CVs.
Due to the hardness of Source L's spectrum and its distance from the core's half-mass radius, Source L is likely a background source. This is roughly consistent with the 2.1 background sources expected from \citet{Giacconi01} extragalactic source counts (see Section \ref{sec:RD}).

In 2010, NGC 6652 was detected by the $\it{Fermi}$ Large Area Telescope, suggesting that the cluster hosts a substantial population of MSPs. \citet{Abdo10} have estimated the number of MSPs in NGC 6652 to be 54$^{+27}_{-25}$; however, previous studies using radio observations have placed this number much lower \citep{Kulkarni90}. A MSP coincident with NGC 6652 has recently been reported; however, it remains unclear whether it is truly associated with the cluster \citep{DeCesar11}.  MSPs could be one or more of the X-ray sources in NGC 6652. Previous observations of MSPs in other globular clusters (e.g. \cite{Bogdanov05, Bogdanov10a}) have shown that the brightest MSPs have hard spectra with luminosities typically $<$10$^{32}$ erg s$^{-1}$. Thus, some of the fainter sources in the cluster may be MSPs. 


\subsection{X-ray Source Population}

NGC 6652 is unusual in hosting two relatively bright X-ray sources. Calculations of the encounter rate, $\Gamma$, for NGC 6652 vary significantly, ranging from 0.2$\%$ to 2.3$\%$ of the Galactic value, depending on whether values are taken from \citet{Harris96} or \citet{Noyola06}. A full analysis of the effects of uncertainties in input parameters on the encounter rate is underway  (Bahramian et al., in preparation). A preliminary estimate of the uncertainty in NGC 6652's encounter rate, using quoted measurements and uncertainty in surface brightness and core radius from \citet{Noyola06}, and extinction and distance measurements from \citet{Harris10}, yields an encounter rate $\Gamma$ =1.5-4\% of the total Galactic globular cluster interaction rate.  Such a high encounter rate would explain NGC 6652's containing two XRBs above $10^{33}$ ergs/s.  

We present cluster parameters and X-ray source numbers in three luminosity bins for several clusters in Table \ref{tab:interactionrates}, ordered by central luminosity density.  
The ratio of the number of brighter to fainter X-ray sources 
(e.g. $>$$10^{33}$ to $10^{32-33}$ erg s$^{-1}$, or $10^{32-33}$ to $10^{31-32}$ erg s$^{-1}$) in NGC 6652, and other clusters with very dense cores,  
is relatively high, indicating a relatively flat luminosity function. 
These data support the argument by \citet{Pooley02b} that clusters with higher central densities have flatter luminosity functions, and the argument by \citet{Heinke03d} that the differences above $10^{31}$ ergs/s are driven primarily by changes within the CV population.  Where detailed optical followup has been done \citep[e.g.][]{Pooley02a,Edmonds03a,Lugger07,Cohn10}, a majority of the X-ray sources that we consider ($L_X$(0.5--2.5 keV)$>10^{31}$ ergs/s) are associated with CVs and qLMXBs.\footnote{This agrees with \citet{Pooley06} who used the 0.5--6 keV band.}  CVs make up the majority of the two lower bins, while the top bin is comprised of accreting NSs (qLMXBs and LMXBs).  So the CVs tend to show higher luminosities on average in the densest clusters; the lack of numerous qLMXBs in NGC 6652 and M15 suggests that NSs are also affected.  Overall, the data suggest that clusters that reach very high central density efficiently convert their numerous lower luminosity X-ray binaries into a smaller group of higher luminosity CVs and LMXBs.

Core collapse may involve recurrent episodes of extremely high central density \citep{Fregeau03}, in which new binaries are formed and older binaries are destroyed or ejected through interactions.  The highest-density clusters are also those clusters where mass-transferring binaries are most rapidly destroyed (or ejected) through further interactions \citep{Verbunt02}.  Binaries formed through interactions will tend to consist of two relatively massive stars.  These will produce relatively high mass transfer rates, and thus relatively high $L_X$ systems.  Among CVs, magnetic channeling of accretion is required to enable high $L_X$ \citep[e.g.][]{Edmonds03b}; for qLMXBs, the quiescent $L_X$ may reflect the time-averaged mass-transfer rate \citep{Brown98}.\footnote{Though neutrino losses from e.g. direct URCA can reduce their quiescent $L_X$, \citealt{Yakovlev04}.}  As systems age, the companion is whittled away, leading to lower mass-transfer rates and thus fainter systems.  In the densest systems like NGC 6652, X-ray binaries may be destroyed or ejected before they can dim significantly.  The extended radial distribution of X-ray binaries in NGC 6652, particularly the large offsets of sources A and B, indeed suggests recent ejection.

\subsection{Future Directions}

Source B's rapid X-ray and optical \citep{Engel12} flaring would be worth studying simultaneously (e.g. with ULTRACAM and \Chandra), to clearly distinguish the details of accretion.  Further \HST\  analysis is needed to confirm (e.g. through star counts) the sharp central cusp and core-collapsed state of NGC 6652 identified through surface brightness studies by \citet{Noyola06}, as well as to identify optical counterparts of the fainter X-ray sources.   
 If source A drops into quiescence during the \Chandra\ era, a deeper study of NGC 6652's X-ray sources could confirm the unusual radial distribution and luminosity function trends we have suggested here.   
Detailed studies of the X-ray binary populations of other clusters of high central density, together with continuing Monte Carlo and N-body simulations, may help us to understand how core collapse proceeds.


\acknowledgements

WSS, COH \& AB acknowledge support by NSERC and by an Alberta Ingenuity New Faculty Award.   
PWdetect has been developed by scientists at Osservatorio Astronomico di Palermo G. S. Vaiana thanks to Italian CNAA and MURST (COFIN) grants.

\begin{deluxetable}{cccccc}
\tablewidth{6truein}
\tablecaption{\textbf{X-ray Sources in NGC 6652}}
\tablehead{
\colhead{\textbf{Source}}  & R.A.& Decl. & Err & Counts & Luminosity  \\
\colhead{} & & & & (0.3--7 keV) & (0.5--10 keV) \\
}
\startdata
A & 18:35:43.671  & -32:59:26.31 & 0.29 & 1267$^{+37}_{-36}$$^\alpha$ & 7.0$^{+0.7}_{-0.7}\times10^{35}$ \\
B & 18:35:44.564 & -32:59:38.49 & 0.28 & 4634$^{+69}_{-68}$ & - \\
C & 18:35:45.752 & -32:59:23.24 & 0.30 & 596$^{+25}_{-24}$ & - \\
D & 18:35:45.651 & -32:59:26.17 & 0.30 & 631$^{+26}_{-25}$& - \\
E & 18:35:46.240 & -32:59:29.36 & 0.33 & 99$^{+11}_{-10}$  & -\\
F & 18:35:47.743 & -32:59:19.76 & 0.36 & 35$^{+7}_{-6}$ & -\\
H & 18:35:45.473 & -32:59:25.48 & 0.40 & 16$^{+5}_{-4}$ & 4.5$^{+1.4}_{-1.1}\times10^{31}$ \\
I  & 18:35:45.818 & -32:59:35.82 & 0.40 & 15$^{+5}_{-4}$ & 4.2$^{+1.4}_{-1.1}\times10^{31}$ \\
J & 18:35:45.178  & -32:59:34.54 & 0.43 & 11$^{+4}_{-3}$ & 3.1$^{+1.1}_{-0.8}\times10^{31}$ \\
K & 18:35:44.078 & -32:59:00.54 & 0.45 & 9$^{+4}_{-3}$ & 2.5$^{+1.1}_{-0.8}\times10^{31}$ \\
L & 18:35:43.891 & -33:00:45.63 & 0.72 & 4$^{+3}_{-2}$ & 1.1$^{+0.8}_{-0.6}\times10^{31}$ \\
\enddata 
\tablecomments{Positions, relative position errors and counts obtained for each observed source in NGC 6652. Values for sources A-F were determined by running the CIAO tool $\it{wavdetect}$ over 0.3--7 keV. Values for sources H-L were determined by running the $\it{PWDetect}$ script over the same energy range. Positional errors are 95\% confidence errors from Equation (5) of \citet{Hong05}.  Values for errors in the counts are indicative of 1$\sigma$ and are derived from \citet{Gehrels86}. Luminosities for sources B-F can be found in Table \ref{tab:specfits}. Luminosities for sources H-L are based upon the number of counts (see text) and are listed in erg s$^{-1}$. $^\alpha$ From readout streak.
 }
\label{tab:positions}
\end{deluxetable}

\begin{deluxetable}{cccccccc}
\tablewidth{7truein}
\tablecaption{\textbf{Count rates for Source B}}
\tablehead{
&  \multicolumn{2} { c }{Counts} & & $N_H$ & $kT$ & $L_X$  \\
\colhead{\textbf{Count Rate}}  & 0.3--1 keV  & 0.3--7 keV  & Hardness & (10$^{20}$ cm$^{-2}$) & keV & (0.5--10 keV) &$\chi^2_{\nu}$/dof \\
}
\startdata
$<$ 0.05 & 71 & 424 & 0.17$\pm$0.02 & 23$^{+11}_{-9}$  & $>$33 & 5.8$^{+0.7}_{-0.7}\times10^{33}$ & 1.05/10 \\
0.05 - 0.075 & 67 & 410 & 0.16$\pm$0.02 & 27$^{+17}_{-15}$ & $>$5 & 1.0$^{+0.2}_{-0.2}\times10^{34}$ & 0.89/9\\
0.075 - 0.1 & 109 & 679 & 0.16$\pm$0.02 & 21$^{+9}_{-8}$ & 16$^{+36}_{-7}$  & 1.4$^{+0.2}_{-0.1}\times10^{34}$ & 0.80/18\\
0.1 - 0.15& 135 & 694 & 0.20$\pm$0.02  & 13$^{+8}_{-7}$ & 12$^{+24}_{-5}$ & 1.9$^{+0.2}_{-0.2}\times10^{34}$ & 0.53/18\\
0.15 - 0.2& 147 & 720 & 0.20$\pm$0.02  & 12$^{+7}_{-6}$ & 7$^{+8}_{-3}$ & 2.4$^{+0.3}_{-0.2}\times10^{34}$ & 0.48/19\\
0.2 - 0.27& 107 & 507 & 0.21$\pm$0.02  & 10$^{+8}_{-8}$ & $>$7 & 3.8$^{+0.7}_{-0.5}\times10^{34}$ & 0.84/12\\
0.27 - 0.35 & 130 & 570 & 0.23$\pm$0.02  & $<$9 & $>$9 & 4.7$^{+0.8}_{-0.8}\times10^{34}$& 1.35/14\\
$>$ 0.35 & 106 & 491 & 0.22$\pm$0.02  & $<$13 & 13$^{+51}_{-7}$ & 6.5$^{+1.1}_{-1.0}\times10^{34}$ & 0.36/12\\
\enddata
\tablecomments{Number of counts observed over indicated energy ranges corresponding to particular count rate ranges. 
Count rate ranges are listed in counts/s.
The hardness ratio is quantified by examining the fraction of counts in the low energy range (0.3--1 keV) against a larger energy range (0.3--7 keV).
Errors in the hardness ratio are indicative of 1$\sigma$ and are derived from \citet{Gehrels86}.
The hydrogen column density ($N_H$), plasma temperature ($kT$) and luminosity are taken from fitting the spectra over each range to an absorbed hot X-ray plasma emission model (XSPEC model $phabs * mekal$).
Luminosity is in erg s$^{-1}$.
The five ranges with the highest count rate include a pileup component. } 
\label{tab:specfitsb}
\end{deluxetable}

\clearpage

\begin{deluxetable}{ccccccc}
\tablewidth{6.7truein}
\tablecaption{\textbf{Spectral Fits}}
\tablehead{
\colhead{Source}  & Model & $N_H$ & $\Gamma$ & kT  & $L_X$ & $\chi^2_{\nu}$/dof  \\
\colhead {} & & (10$^{20}$ cm$^{-2}$) & & (keV) & (0.5--10 keV) & \\
}
\startdata

\textbf{B} & \textbf{POW} & 11$^{+3}_{-3}$ & 1.3$^{+0.1}_{-0.1}$ & - & 1.7$^{+0.1}_{-0.1}\times10^{34}$ & 1.20/39   \\ 
\textbf{B} & \textbf{MEKAL} & 10$^{+2}_{-2}$ & - & 34$^{+28}_{-13}$ & 1.6$^{+0.1}_{-0.1}\times10^{34}$ & 1.13/39   \\
 \hline \\
\textbf{C} & \textbf{POW} & (5) & 1.6 & - & 1.3$\times10^{33}$ & 4.18/14   \\ 
\textbf{C} & \textbf{MEKAL+MEKAL} & $<$32 & - & $>$17 & 1.4$^{+0.2}_{-0.2}\times10^{33}$ & 1.06/11   \\
\textbf{-} & \textbf{2$^{nd}$ MEKAL} & - & - & 0.18$^{+0.03}_{-0.03}$ & 3.2$^{+0.8}_{-1.9}\times10^{32}$ & -   \\
\textbf{C} & \textbf{MEKAL+BBODY} & $<$19 & - & $>$18 & 1.4$^{+0.2}_{-0.2}\times10^{33}$ & 1.05/11   \\
\textbf{-} & \textbf{BBODY} & - & - & 0.07$^{+0.01}_{-0.02}$ & 2.3$^{+6.6}_{-1.0}\times10^{32}$ & -   \\
 \hline \\
\textbf{D} & \textbf{POW} & 42$^{+14}_{-19} $& 5.1$^{+0.8}_{-1.2}$ & - & 4.1$^{+6.1}_{-2.1}\times10^{33}$ & 0.77/14   \\ 
\textbf{D} & \textbf{MEKAL} & (5) & - & 1.3 & 8.3$\times10^{32}$ & 3.52/15   \\
\textbf{D} & \textbf{NSATMOS} & 9$^{+2}_{-2}$ & - & 0.115 $^{+0.002}_{-0.002}$ & 9.0$^{+4.0}_{-2.1}\times10^{32}$ & 0.65/15   \\
 \hline \\
\textbf{E} & \textbf{POW} & 53$^{+48}_{-36} $& 2.2$^{+1.1}_{-0.8}$ & - & 3.8$^{+3.8}_{-1.1}\times10^{32}$ & 0.62/6   \\ 
\textbf{E} & \textbf{MEKAL} & 34$^{+35}_{-25} $& - & $>$1.8 & 2.9$^{+0.9}_{-0.7}\times10^{32}$ & 0.78/6   \\
 \hline \\ 
 \textbf{F} & \textbf{POW} & $<$35 & 1.7$^{+1.1}_{-0.5}$ & - & 1.1$^{+0.8}_{-0.5}\times10^{32}$ & 1.03/1   \\ 
\textbf{F} & \textbf{MEKAL} & $<$82 & - & $>$1.9 & 1.4$^{+0.6}_{-0.5}\times10^{32}$ & 1.11/1   \\
\textbf{F} & \textbf{NSATMOS} & (5) & - & 8.5$ \times10^{-2}$ & 1.3$\times10^{32}$ & 6.54/2   \\

\enddata
\tablecomments{ Spectral fits for X-ray sources in NGC 6652. 
Sources B through E are fit from 0.5--10 keV; however, due to a low number of counts (42) Source F is fit from 0--10 keV.
All models included absorption (XSPEC model $phabs$).
For poorly fit models ($\chi^2_{\nu} >$ 2) $N_H$ was kept at the cluster value and errors were not calculated.
The luminosity stated is for the unabsorbed X-ray luminosity in erg s$^{-1}$. 
Two component models are continued on a second line and give the X-ray luminosity separated by components.
The total unabsorbed 0.5--10 keV luminosities for Source C from the \textbf{MEKAL+MEKAL} model and the \textbf{MEKAL+BBODY} model are respectively 1.8$^{+0.1}_{-0.2}\times10^{33}$ erg s$^{-1}$ and 1.6$^{+0.7}_{-0.2}\times10^{33}$ erg s$^{-1}$.
The final column lists the reduced chi-squared and the degrees of freedom for each model. }
\label{tab:specfits}
\end{deluxetable}

\clearpage

\begin{deluxetable}{ccccccccc}
\tablewidth{7truein}
\tablecaption{\textbf{Parameters and X-ray Sources of Selected Clusters}}
\tablehead{
\colhead{Cluster } & Distance & $\rho_0$ & $r_c$ & $\Gamma$ &\multicolumn{3} { c }{Number of X-ray Sources} & References  \\
\colhead{ }  & (kpc) & ($\Lsun$ pc$^{-3}$) & (pc) & (\% Gal.) & ($>10^{33}$) &  ($10^{32-33}$) &  ($10^{31-32}$) &   \\
}
\startdata

\textbf{47 Tuc} 	& 4.5  & $7.8\times10^4$ & 0.46  & 2.9   & 1 & 3   & 27 & 1 \\
\textbf{M28}           & 5.5  & $1.3\times10^5$ & 0.26  & 2.0   & 1 & 2   & 14 &   2,3\\
\textbf{M80}           &10.0 & $1.4\times10^5$ & 0.22  & 1.6   & 0 & 4   & 12 & 4\\
\textbf{NGC 6440} & 8.5  & $1.7\times10^5$ & 0.35   & 3.6   & 1 & 10 & $>$13 & 5,3 \\
\textbf{NGC 6388} & 9.9  & $2.3\times10^5$ & 0.21   & 3.1   & 0  & 13 & 27 & 6\\
\textbf{NGC 6752} & 4.0  & $2.3\times10^5$ & 0.13   & 1.2   & 0 & 1   & 8   & 7\\
\textbf{M30}           & 8.1  & $3.6\times10^5$ & 0.063 & 0.6   & 1 & 0   & 3  & 8\\
\textbf{NGC 6397} & 2.3  & $4.1\times10^5$ & 0.041 & 0.3   & 0 & 2   & 7   & 9\\
\textbf{NGC 6652} &10.0 & $1.0\times10^6$ & 0.058 & 2.3   & 2 & 3   & 4   & 10\\
\textbf{M15}           &10.4 & $3.5\times10^6$ & 0.050 & 6.7 & 3 & 2   & $>$2   & 11,12\\

\enddata
\tablecomments{ Parameters and X-ray source numbers from clusters with numerous published X-ray sources above $L_X=10^{31}$ ergs/s (Terzan 5 omitted due to uncertainty in parameters). 
The numbers of X-ray sources include only sources within the cluster half-mass radius, using luminosities in the 0.5--2.5 keV range from the deepest observation (or average published luminosities for M15 sources).  M15 and NGC 6440 are incomplete in the lowest bin.
Central luminosity densities (for the core-collapsed, we average within the core radius) and core radii are calculated using values from \citet{Noyola06} where given, while other values are from \citet{Harris10}. 
Close encounter rates $\Gamma \propto \rho_0^{1.5}r_c^2$ are given as a percentage of the total Galactic globular cluster system rate.
References.-
(1) \citet{Heinke05}; 
(2) \citet{Becker03};
(3) \citet{Heinke03d}; 
(4) \citet{Heinke03c}; 
(5) \citet{Pooley02b}; 
(6) \citet{Maxwell12};
(7) \citet{Pooley02a};
(8) \citet{Lugger07}; 
 (9) \citet{Grindlay01b}; 
(10) this work; 
(11) \citet{Hannikainen05};
 (12) \citet{Heinke09}. 
}
\label{tab:interactionrates}
\end{deluxetable}


\bibliographystyle{apj}

\bibliography{NGC6652bib}

\begin{thebibliography}{70}
\expandafter\ifx\csname natexlab\endcsname\relax\def\natexlab#1{#1}\fi

\bibitem[{{Abdo} {et~al.}(2010){Abdo}, {Ackermann}, {Ajello}, \& {et
  al.}}]{Abdo10}
{Abdo}, A.~A. {et al.} 2010, \aap, 524, A75

\bibitem[{{Baskill} {et~al.}(2005){Baskill}, {Wheatley}, \&
  {Osborne}}]{Baskill05}
{Baskill}, D.~S., {Wheatley}, P.~J., \& {Osborne}, J.~P. 2005, \mnras, 357, 626

\bibitem[{{Becker} {et~al.}(2003){Becker}, {Swartz}, {Pavlov}, {Elsner},
  {Grindlay}, {Mignani}, {Tennant}, {Backer}, {Pulone}, {Testa}, \&
  {Weisskopf}}]{Becker03}
{Becker}, W. {et al.} 2003, \apj, 594, 798

\bibitem[{{Bogdanov} {et~al.}(2005){Bogdanov}, {Grindlay}, \& {van den
  Berg}}]{Bogdanov05}
{Bogdanov}, S., {Grindlay}, J.~E., \& {van den Berg}, M. 2005, \apj, 630, 1029

\bibitem[{{Bogdanov} {et~al.}(2010){Bogdanov}, {van den Berg}, {Heinke},
  {Cohn}, {Lugger}, \& {Grindlay}}]{Bogdanov10a}
{Bogdanov}, S., {van den Berg}, M., {Heinke}, C.~O., {Cohn}, H.~N., {Lugger},
  P.~M., \& {Grindlay}, J.~E. 2010, \apj, 709, 241

\bibitem[{{Brown} {et~al.}(1998){Brown}, {Bildsten}, \& {Rutledge}}]{Brown98}
{Brown}, E.~F., {Bildsten}, L., \& {Rutledge}, R.~E. 1998, \apjl, 504, L95

\bibitem[{{Cash}(1979)}]{Cash79}
{Cash}, W. 1979, \apj, 228, 939

\bibitem[{{Cohn} {et~al.}(2010){Cohn}, {Lugger}, {Couch}, {Anderson}, {Cool},
  {van den Berg}, {Bogdanov}, {Heinke}, \& {Grindlay}}]{Cohn10}
{Cohn}, H.~N. {et al.}  2010, \apj, 722, 20

\bibitem[{{Coomber} {et~al.}(2011){Coomber}, {Heinke}, {Cohn}, {Lugger}, \&
  {Grindlay}}]{Coomber11}
{Coomber}, G., {Heinke}, C.~O., {Cohn}, H.~N., {Lugger}, P.~M., \& {Grindlay},
  J.~E. 2011, \apj, 735, 95

\bibitem[{{Damiani} {et~al.}(1997{\natexlab{a}}){Damiani}, {Maggio}, {Micela},
  \& {Sciortino}}]{Damiani97i}
{Damiani}, F., {Maggio}, A., {Micela}, G., \& {Sciortino}, S.
  1997{\natexlab{a}}, \apj, 483, 350

\bibitem[{{Damiani} {et~al.}(1997{\natexlab{b}}){Damiani}, {Maggio}, {Micela},
  \& {Sciortino}}]{Damiani97ii}
---. 1997{\natexlab{b}}, \apj, 483, 370

\bibitem[{{D'Angelo} \& {Spruit}(2010)}]{D'Angelo10}
{D'Angelo}, C.~R. \& {Spruit}, H.~C. 2010, \mnras, 406, 1208

\bibitem[{{Davis}(2001)}]{Davis01}
{Davis}, J.~E. 2001, \apj, 562, 575

\bibitem[{{DeCesar} {et~al.}(2011){DeCesar}, {Ransom}, \& {Ray}}]{DeCesar11}
{DeCesar}, M.~E., {Ransom}, S.~M., \& {Ray}, P.~S. 2011, arXiv1111.0365

\bibitem[{{Del Santo} {et~al.}(2007){Del Santo}, {Sidoli}, {Mereghetti},
  {Bazzano}, {Tarana}, \& {Ubertini}}]{delSanto07}
{Del Santo}, M., {Sidoli}, L., {Mereghetti}, S., {Bazzano}, A., {Tarana}, A.,
  \& {Ubertini}, P. 2007, \aap, 468, L17

\bibitem[{{Deutsch} {et~al.}(1998){Deutsch}, {Margon}, \&
  {Anderson}}]{Deutsch98}
{Deutsch}, E.~W., {Margon}, B., \& {Anderson}, S.~F. 1998, \aj, 116, 1301

\bibitem[{{Edmonds} {et~al.}(2003{\natexlab{a}}){Edmonds}, {Gilliland},
  {Heinke}, \& {Grindlay}}]{Edmonds03a}
{Edmonds}, P.~D., {Gilliland}, R.~L., {Heinke}, C.~O., \& {Grindlay}, J.~E.
  2003{\natexlab{a}}, \apj, 596, 1177

\bibitem[{{Edmonds} {et~al.}(2003{\natexlab{b}}){Edmonds}, {Gilliland},
  {Heinke}, \& {Grindlay}}]{Edmonds03b}
---. 2003{\natexlab{b}}, \apj, 596, 1197

\bibitem[{{Edmonds} {et~al.}(1999){Edmonds}, {Grindlay}, {Cool}, {Cohn},
  {Lugger}, \& {Bailyn}}]{Edmonds99}
{Edmonds}, P.~D., {Grindlay}, J.~E., {Cool}, A., {Cohn}, H., {Lugger}, P., \&
  {Bailyn}, C. 1999, \apj, 516, 250

\bibitem[{{Engel} {et~al.}(2012){Engel}, {Heinke}, {Sivakoff}, {El-Shamouty},
  \& {Edmonds}}]{Engel12}
{Engel}, M.~C., {Heinke}, C.~O., {Sivakoff}, G.~R., {El-Shamouty}, K.~G., \&
  {Edmonds}, P.~D. 2012, \apj, 747, 119

\bibitem[{{Fregeau} {et~al.}(2003){Fregeau}, {G{\" u}rkan}, {Joshi}, \&
  {Rasio}}]{Fregeau03}
{Fregeau}, J.~M., {G{\" u}rkan}, M.~A., {Joshi}, K.~J., \& {Rasio}, F.~A. 2003,
  \apj, 593, 772

\bibitem[{{Gehrels}(1986)}]{Gehrels86}
{Gehrels}, N. 1986, \apj, 303, 336

\bibitem[{{Giacconi} {et~al.}(2001){Giacconi}, {Rosati}, {Tozzi}, \& {et
  al.}}]{Giacconi01}
{Giacconi}, R. {et al.} 2001, \apj, 551, 624

\bibitem[{{Gregory} \& {Loredo}(1992)}]{Gregory92}
{Gregory}, P.~C. \& {Loredo}, T.~J. 1992, \apj, 398, 146

\bibitem[{{Grindlay} {et~al.}(2002){Grindlay}, {Camilo}, {Heinke}, {Edmonds},
  {Cohn}, \& {Lugger}}]{Grindlay02}
{Grindlay}, J.~E., {Camilo}, F., {Heinke}, C.~O., {Edmonds}, P.~D., {Cohn}, H.,
  \& {Lugger}, P. 2002, \apj, 581, 470

\bibitem[{{Grindlay} {et~al.}(1995){Grindlay}, {Cool}, {Callanan}, {Bailyn},
  {Cohn}, \& {Lugger}}]{Grindlay95}
{Grindlay}, J.~E., {Cool}, A.~M., {Callanan}, P.~J., {Bailyn}, C.~D., {Cohn},
  H.~N., \& {Lugger}, P.~M. 1995, \apjl, 455, L47

\bibitem[{{Grindlay} {et~al.}(2001{\natexlab{a}}){Grindlay}, {Heinke},
  {Edmonds}, \& {Murray}}]{Grindlay01}
{Grindlay}, J.~E., {Heinke}, C., {Edmonds}, P.~D., \& {Murray}, S.~S.
  2001{\natexlab{a}}, Science, 292, 2290

\bibitem[{{Grindlay} {et~al.}(2001{\natexlab{b}}){Grindlay}, {Heinke},
  {Edmonds}, {Murray}, \& {Cool}}]{Grindlay01b}
{Grindlay}, J.~E., {Heinke}, C.~O., {Edmonds}, P.~D., {Murray}, S.~S., \&
  {Cool}, A.~M. 2001{\natexlab{b}}, \apjl, 563, L53

\bibitem[{{Hannikainen} {et~al.}(2005){Hannikainen}, {Charles}, {van Zyl},
  {Kong}, {Homer}, {Hakala}, {Naylor}, \& {Davies}}]{Hannikainen05}
{Hannikainen}, D.~C., {Charles}, P.~A., {van Zyl}, L., {Kong}, A.~K.~H.,
  {Homer}, L., {Hakala}, P., {Naylor}, T., \& {Davies}, M.~B. 2005, \mnras,
  357, 325

\bibitem[{{Harris}(1996)}]{Harris96}
{Harris}, W.~E. 1996, \aj, 112, 1487

\bibitem[{{Harris}(2010)}]{Harris10}
---. 2010, ArXiv e-prints

\bibitem[{{Heinke} {et~al.}(2009){Heinke}, {Cohn}, \&
  {Lugger}}]{Heinke09}
{Heinke}, C.~O., {Cohn}, H.~N., \& {Lugger}, P.~M. 2009, \apj,
  692, 584

\bibitem[{{Heinke} {et~al.}(2001){Heinke}, {Edmonds}, \& {Grindlay}}]{Heinke01}
{Heinke}, C.~O., {Edmonds}, P.~D., \& {Grindlay}, J.~E. 2001, \apj, 562, 363

\bibitem[{{Heinke} {et~al.}(2005){Heinke}, {Grindlay}, {Edmonds},
  {Cohn}, {Lugger}, {Camilo}, {Bogdanov}, \& {Freire}}]{Heinke05}
{Heinke}, C.~O. {et al.} 2005, \apj, 625, 796

\bibitem[{{Heinke} {et~al.}(2003{\natexlab{a}}){Heinke}, {Grindlay}, {Edmonds},
  {Lloyd}, {Murray}, {Cohn}, \& {Lugger}}]{Heinke03c}
{Heinke}, C.~O., {Grindlay}, J.~E., {Edmonds}, P.~D., {Lloyd}, D.~A., {Murray},
  S.~S., {Cohn}, H.~N., \& {Lugger}, P.~M. 2003{\natexlab{a}}, \apj, 598, 516

\bibitem[{{Heinke} {et~al.}(2003{\natexlab{b}}){Heinke}, {Grindlay}, {Lugger},
  {Cohn}, {Edmonds}, {Lloyd}, \& {Cool}}]{Heinke03d}
{Heinke}, C.~O., {Grindlay}, J.~E., {Lugger}, P.~M., {Cohn}, H.~N., {Edmonds},
  P.~D., {Lloyd}, D.~A., \& {Cool}, A.~M. 2003{\natexlab{b}}, \apj, 598, 501

\bibitem[{{Heinke} {et~al.}(2006){Heinke}, {Wijnands}, {Cohn}, {Lugger},
  {Grindlay}, {Pooley}, \& {Lewin}}]{Heinke06b}
{Heinke}, C.~O., {Wijnands}, R., {Cohn}, H.~N., {Lugger}, P.~M., {Grindlay},
  J.~E., {Pooley}, D., \& {Lewin}, W.~H.~G. 2006, ApJ, 651, 1098

\bibitem[{{Hertz} \& {Grindlay}(1983)}]{Hertz83}
{Hertz}, P. \& {Grindlay}, J.~E. 1983, \apj, 275, 105

\bibitem[{{Hong} {et~al.}(2005){Hong},{van den Berg}, {Schlegel}, {Grindlay}, {Koenig}, {Laycock}, \& {Zhao}}]{Hong05}
{Hong}, J., {van den Berg}, M., {Schlegel}, E.~M., {Grindlay}, J.~E., {Koenig}, X., {Laycock}, S. \& {Zhao}, P. 2005, \apj, 635, 907

\bibitem[{{Illarionov} \& {Sunyaev}(1975)}]{Illarionov75}
{Illarionov}, A.~F. \& {Sunyaev}, R.~A. 1975, \aap, 39, 185

\bibitem[{{Ivanova} {et~al.}(2006){Ivanova}, {Heinke}, {Rasio}, {Taam},
  {Belczynski}, \& {Fregeau}}]{Ivanova06}
{Ivanova}, N., {Heinke}, C.~O., {Rasio}, F.~A., {Taam}, R.~E., {Belczynski},
  K., \& {Fregeau}, J. 2006, \mnras, 372, 1043

\bibitem[{{King}(2000)}]{King00}
{King}, A.~R. 2000, \mnras, 315, L33

\bibitem[{{Kulkarni} {et~al.}(1990){Kulkarni}, {Narayan}, \&
  {Romani}}]{Kulkarni90}
{Kulkarni}, S.~R., {Narayan}, R., \& {Romani}, R.~W. 1990, \apj, 356, 174

\bibitem[{{Lugger} {et~al.}(2007){Lugger}, {Cohn}, {Heinke}, {Grindlay}, \&
  {Edmonds}}]{Lugger07}
{Lugger}, P.~M., {Cohn}, H.~N., {Heinke}, C.~O., {Grindlay}, J.~E., \&
  {Edmonds}, P.~D. 2007, \apj, 657, 286

\bibitem[{{Maxwell} {et~al.}(2012){Maxwell}, {Lugger}, {Cohn}, {Grindlay},
  {Heinke}, {Budac}, {Drukier}, \& {Bailyn}}]{Maxwell12}
{Maxwell}, T.~E. {et al.} 2012, ApJ,
  submitted

\bibitem[{{McLaughlin} \& {van der Marel}(2005)}]{McLaughlin05}
{McLaughlin}, D.~E. \& {van der Marel}, R.~P. 2005, \apjs, 161, 304

\bibitem[{{Mukai} \& {Smale}(2000)}]{Mukai00}
{Mukai}, K. \& {Smale}, A., 2000, \apj, 533, 352

\bibitem[{{Muno} {et~al.}(2005){Muno}, {Lu}, {Baganoff}, {Brandt}, {Garmire},
  {Ghez}, {Hornstein}, \& {Morris}}]{Muno05b}
{Muno}, M.~P. {et al.} 2005, \apj, 633,
  228

\bibitem[{{Noyola} \& {Gebhardt}(2006)}]{Noyola06}
{Noyola}, E. \& {Gebhardt}, K. 2006, \aj, 132, 447

\bibitem[{{Parmar} {et~al.}(2001){Parmar}, {Oosterbroek}, {Sidoli}, {Stella},
  \& {Frontera}}]{Parmar01}
{Parmar}, A.~N., {Oosterbroek}, T., {Sidoli}, L., {Stella}, L., \& {Frontera},
  F. 2001, \aap, 380, 490

\bibitem[{{Patruno} {et~al.}(2009){Patruno}, {Watts}, {Klein-Wolt}, {Wijnands},
  \& {van der Klis}}]{Patruno09}
{Patruno}, A., {Watts}, A.~L., {Klein-Wolt}, M., {Wijnands}, R., \& {van der
  Klis}, M. 2009, \apj, 707, 1296

\bibitem[{{Patterson} \& {Raymond}(1985)}]{Patterson85}
{Patterson}, J. \& {Raymond}, J.~C. 1985, \apj, 292, 535

\bibitem[{{Pooley} \& {Hut}(2006)}]{Pooley06}
{Pooley}, D. \& {Hut}, P. 2006, \apjl, 646, L143

\bibitem[{{Pooley} {et~al.}(2002{\natexlab{a}}){Pooley}, {Lewin}, {Homer}, \&
  {et al.}}]{Pooley02a}
{Pooley}, D. {et al.} 2002{\natexlab{a}},
  \apj, 569, 405

\bibitem[{{Pooley} {et~al.}(2002{\natexlab{b}}){Pooley}, {Lewin}, {Verbunt}, \&
  {et al.}}]{Pooley02b}
{Pooley}, D. {et al.} 2002{\natexlab{b}},
  \apj, 573, 184

\bibitem[{{Ramsay} {et~al.}(2004){Ramsay}, {Cropper}, {Wu}, {Mason},
  {C{\'o}rdova}, \& {Priedhorsky}}]{Ramsay04}
{Ramsay}, G., {Cropper}, M., {Wu}, K., {Mason}, K.~O., {C{\'o}rdova}, F.~A., \&
  {Priedhorsky}, W. 2004, \mnras, 350, 1373

\bibitem[{{Romanova} {et~al.}(2002){Romanova}, {Ustyugova}, {Koldoba}, \&
  {Lovelace}}]{Romanova02}
{Romanova}, M.~M., {Ustyugova}, G.~V., {Koldoba}, A.~V., \& {Lovelace},
  R.~V.~E. 2002, \apj, 578, 420

\bibitem[{{Romanova} {et~al.}(2005){Romanova}, {Ustyugova}, {Koldoba}, \&
  {Lovelace}}]{Romanova05}
---. 2005, \apjl, 635, L165

\bibitem[{{Romanova} {et~al.}(2011){Romanova}, {Ustyugova}, {Koldoba}, \&
  {Lovelace}}]{Romanova11}
---. 2011, \mnras, 416, 416

\bibitem[{{Shara} {et~al.}(1996){Shara}, {Bergeron}, {Gilliland}, {Saha}, \&
  {Petro}}]{Shara96}
{Shara}, M.~M., {Bergeron}, L.~E., {Gilliland}, R.~L., {Saha}, A., \& {Petro},
  L. 1996, \apj, 471, 804

\bibitem[{{Sidoli} {et~al.}(2001){Sidoli},{Parmar}, {Oosterbroek}, {Stella}, {Verbunt}, {Masetti}, \& {Dal Fiume}}]{Sidoli01}
{Sidoli}, L., {Parmar}, A.~N., {Oosterbroek}, T., {Stella}, L., {Verbunt}, F., {Masetti}, N., \& {Dal Fiume}, D. 2001, \aap, 368, 451

\bibitem[{{Sidoli} {et~al.}(2008){Sidoli}, {La Palombara}, {Oosterbroek}, \& {Parmar}}]{Sidoli08}
{Sidoli}, L., {La Palombara}, N., {Oosterbroek}, T., \& {Parmar}, A.~N. 2008, \aap, 488, 249


\bibitem[{{Spruit} \& {Taam}(1993)}]{Spruit93}
{Spruit}, H.~C. \& {Taam}, R.~E. 1993, \apj, 402, 593

\bibitem[{{Stacey} {et~al.}(2011){Stacey}, {Heinke}, {Elsner}, {Edmonds},
  {Weisskopf}, \& {Grindlay}}]{Stacey11}
{Stacey}, W.~S., {Heinke}, C.~O., {Elsner}, R.~F., {Edmonds}, P.~D.,
  {Weisskopf}, M.~C., \& {Grindlay}, J.~E. 2011, \apj, 732, 46

\bibitem[{{Swank} \& {Markwardt}(2001)}]{Swank01}
{Swank}, J. \& {Markwardt}, K. 2001, in ASP Conf. Ser. 251, New Century of
  X-ray Astronomy, eds. H. Inoue \& H. Kunieda (San Francisco: ASP), 94

\bibitem[{{Tarana} {et~al.}(2007){Tarana}, {Bazzano}, {Ubertini}, \&
  {Federici}}]{Tarana07}
{Tarana}, A., {Bazzano}, A., {Ubertini}, P., \& {Federici}, M. 2007, in ESA
  Special Publication, Vol. 622, ESA Special Publication, 437+

\bibitem[{{Ustyugova} {et~al.}(2006){Ustyugova}, {Koldoba}, {Romanova}, \&
  {Lovelace}}]{Ustyugova06}
{Ustyugova}, G.~V., {Koldoba}, A.~V., {Romanova}, M.~M., \& {Lovelace},
  R.~V.~E. 2006, \apj, 646, 304

\bibitem[{{Verbunt}(2003)}]{Verbunt02}
{Verbunt}, F. 2003, in ASP Conf. Ser. 296: New Horizons in Globular Cluster
  Astronomy, 245, astro--ph/0210057

\bibitem[{{Verbunt} {et~al.}(1984){Verbunt}, {Elson}, \& {van
  Paradijs}}]{Verbunt84}
{Verbunt}, F., {Elson}, R., \& {van Paradijs}, J. 1984, \mnras, 210, 899

\bibitem[{{Wijnands} {et~al.}(2006){Wijnands}, {in't Zand}, {Rupen},
  {Maccarone}, {Homan}, {Cornelisse}, {Fender}, {Grindlay}, {van der Klis},
  {Kuulkers}, {Markwardt}, {Miller-Jones}, \& {Wang}}]{Wijnands06}
{Wijnands}, R. {et al.} 2006, \aap, 449, 1117

\bibitem[{{Yakovlev} \& {Pethick}(2004)}]{Yakovlev04}
{Yakovlev}, D.~G. \& {Pethick}, C.~J. 2004, \araa, 42, 169

\end{thebibliography}

\end{document}